\documentclass[conference,a4paper]{IEEEtran}
\usepackage{cite}
\usepackage{amsmath,amssymb,amsfonts}
\usepackage{algorithmic}
\usepackage{graphicx}
\usepackage{caption}
\usepackage[british]{babel} 
\usepackage{multirow}
\usepackage{subfig}
\usepackage{tikz}
\usepackage[linesnumbered,ruled,vlined]{algorithm2e}
\usepackage[labelfont=bf]{caption}
\usepackage{textcomp}
\usepackage{xcolor}
\usepackage{balance}
\usepackage{hyperref}
\def\BibTeX{{\rm B\kern-.05em{\sc i\kern-.025em b}\kern-.08em
    T\kern-.1667em\lower.7ex\hbox{E}\kern-.125emX}}
		
\makeatletter
\def\ps@IEEEtitlepagestyle{%
	\def\@oddfoot{\mycopyrightnotice}%
	\def\@oddhead{\hbox{}\@IEEEheaderstyle\leftmark\hfil\thepage}\relax
	\def\@evenhead{\@IEEEheaderstyle\thepage\hfil\leftmark\hbox{}}\relax
	\def\@evenfoot{}%
}
\def\mycopyrightnotice{%
	\begin{minipage}{\textwidth}
		\scriptsize
		\copyright 2021 IEEE.  Personal use of this material is permitted. Permission from IEEE must be obtained for all other uses, in any current or future media, including reprinting/republishing this material for advertising or promotional purposes, creating new collective works, for resale or redistribution to servers or lists, or reuse of any copyrighted component of this work in other works. DOI: \url{https://doi.org/10.1109/VCIP53242.2021.9675326}
	\end{minipage}
}
\makeatother

\SetArgSty{textup}
\SetAlFnt{\scriptsize}

\SetCommentSty{mycommfont}

\def\Figu#1{{\bf Figure \ref{#1}}}

\def\Tabl#1{{\bf Table \ref{#1}}}
\def\Algo#1{{\bf Algorithm \ref{#1}}}

\begin{document}
\setlength{\textfloatsep}{1\baselineskip plus 0.2\baselineskip minus 0.2\baselineskip}
\setlength{\intextsep}{1\baselineskip plus 0.2\baselineskip minus 0.2\baselineskip}
\title{Optimization of Probability Distributions for Residual Coding of Screen Content}

\author{\IEEEauthorblockN{Hannah Och\IEEEauthorrefmark{1}, Tilo Strutz\IEEEauthorrefmark{2}, and Andr\'{e} Kaup\IEEEauthorrefmark{1}}
\IEEEauthorblockA{\IEEEauthorrefmark{1}
Friedrich-Alexander University Erlangen-Nürnberg (FAU)\\Multimedia Communications and Signal Processing, Cauerstr. 7, 91058  Erlangen, Germany}
\IEEEauthorblockA{\IEEEauthorrefmark{2}
Deutsche Telekom AG, Leipzig University of Telecommunications\\
Institute of Communications Engineering, Gustav-Freytag-Str. 43–45, 04277 Leipzig, Germany}
}

\maketitle
\begin{abstract}
Probability distribution modeling is the basis for most competitive methods for lossless coding of screen content. One such state-of-the-art method is known as soft context formation (SCF). For each pixel to be encoded, a probability distribution is estimated based on the neighboring pattern and the occurrence of that pattern in the already encoded image. Using an arithmetic coder, the pixel color can thus be encoded very efficiently, provided that the current color has been observed before in association with a similar pattern. If this is not the case, the color is instead encoded using a color palette or, if it is still unknown, via residual coding. Both palette-based coding and residual coding have significantly worse compression efficiency than coding based on soft context formation.
In this paper, the residual coding stage is improved by adaptively trimming the probability distributions for the residual error. Furthermore, an enhanced probability modeling for indicating a new color depending on the occurrence of new colors in the neighborhood is proposed. These modifications result in a bitrate reduction of up to $\mathbf{2.9\,\%}$ on average. Compared to HEVC (HM-16.21 + SCM-8.8) and FLIF, the improved SCF method saves on average about $\mathbf{11\,\%}$ and $\mathbf{18\,\%}$ rate, respectively. 
\end{abstract}
	%
\section{Introduction}
	%
Images and videos as they can typically be seen on computer screens during office work or the like are called screen content. Such data usually consists of text, computer-generated graphics and animations, but often also camera-captured or highly natural seeming computer-animated content. Generally, screen content consists of repeating patterns and contains a significantly smaller amount of different colors than camera-captured content. 

Coding of screen content is gaining importance with the increasing use of remote desktop applications, online learning, and video conferencing. Unfortunately, conventional compression schemes which are optimized for camera-captured content are often unable to efficiently compress screen content. An indication of the recognition of this issue and a trigger for further research was the incorporation of a screen content coding (SCC) extension in the HEVC standard \cite{Sul12,Xu16}, which integrates dedicated encoding tools for screen content, such as palette coding \cite{Wei16}, adaptive color transform \cite{Li16}, or intra block copy \cite{Xu16b}. With some adaptations, these tools are also incorporated in the following VVC standard \cite{VVCSCC}.

There are, however, few publications pertaining lossless screen content coding. For the HEVC lossless mode, a DPCM-based edge prediction  for screen content is proposed in \cite{San16}. \cite{Kam20} presents pixel-wise blending of sub-predictors for the improvement of lossless intra-prediction. A hybrid adaptive discrete wavelet transform and prediction scheme for JPEG 2000 \cite{JPEG2000} is introduced by \cite{Sta20}. Other competitive methods are based on ideal entropy coding and probability distribution modeling. FLIF (free lossless image format) estimates probabilities on a bitwise level and is designed for both camera-captured and screen content \cite{Sne16}. \cite{Wei13} proposes a parallelized scheme for lossless compression of  medical data using pixel-wise prediction and arithmetic coding. An autoregressive pixel-prediction scheme implementing histogram sparsification for the case of non-natural images is presented in \cite{WeiAm16}.

Soft Context Formation (SCF) \cite{Str19} is a lossless image coding method for screen content that has superior results for images with few colors ($<8000$) with bitrate savings of up to $33\,\%$ compared to HEVC including its SCC extension. It is a pixel-wise method based on arithmetic coding, where probability distributions for colors are estimated based on contexts. However, if the color of a new pixel at a given context is not in the modeled distribution, first palette-based coding is attempted, and if even that is not possible since the color to be encoded is not yet known, residual encoding is performed. Both palette-based coding and residual coding are by far not as efficient as coding based on the context model. As a result, the efficiency of the soft context formation method decreases when images with more colors and more camera captured regions need to be encoded.

In this paper, two modifications to the method of soft context formation are proposed, which aim to improve the coding efficiency of the SCF method for those pixels which cannot be coded using the context model coding. In Section \ref{review}, a short overview of the soft context formation is given, followed by the proposed adaptations in Sections \ref{stage3} and \ref{esc2}. Section \ref{results} presents evaluation results, amongst others a comparison to state-of-the-art codecs, such as the HEVC.
   %
\section{Review of Soft Context Formation}
\label{review}
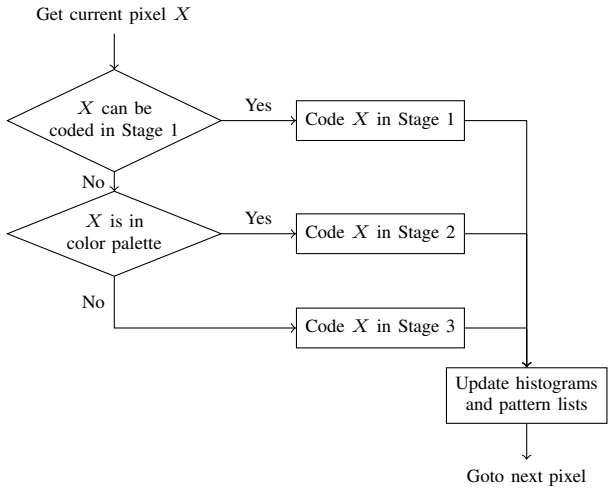
\begin{figure}
	\hfil\usetikzlibrary{shapes.geometric}
\usetikzlibrary{positioning}

\tikzstyle{block} = [draw, fill=white, rectangle, minimum height=1.5em, minimum width=5em, align=center]
\tikzstyle{decision} = [draw, fill=white,diamond, aspect=2,minimum height=2em, minimum width=8em,align=center, inner sep=0pt, outer sep=0pt]
\tikzstyle{input} = [coordinate]
\tikzstyle{output} = [coordinate]
\tikzstyle{pinstyle} = [pin edge={to-,thin,black}]
\tikzstyle{pinstyle2} = [pin edge={-to,thin,black}]

\tikzset{font=\scriptsize}

\begin{tikzpicture}
	\noindent
    \node [decision, pin={[pinstyle]above:Get current pixel $X$}, node distance=1.5cm] (stage1decision){$X$ can be\\ coded in Stage 1};
		\node [decision,below of=stage1decision, node distance=1.5cm] (stage2decision){$X$ is in\\ color palette};
		\node [block, right of=stage1decision, node distance=3.5cm] (stage1){Code $X$ in Stage 1};
		\node [block, right of=stage2decision, node distance=3.5cm] (stage2){Code $X$ in Stage 2};
		\node [block, below of=stage2, node distance=1.25cm] (stage3){Code $X$ in Stage 3};
		\node [block, below right=0.25cm and -0.25cm of stage3,pin={[pinstyle2]below:Goto next pixel}] (update){Update histograms\\and pattern lists};
		
		\draw  [->] (stage1decision) -- node [name=Yes1, midway, above] {Yes} (stage1);
		\draw  [->] (stage2decision) -- node [name=Yes2, midway, above] {Yes} (stage2);
		\draw  [->] (stage1decision) -- node [name=No1, midway, left] {No} (stage2decision);
		\draw [->] (stage2decision) |- node [name=No2, near start, left] {No} (stage3);
		\draw [->] (stage1) -| (update);
		\draw [->] (stage2) -| (update);
		\draw [->] (stage3) -| (update);
\end{tikzpicture}
		\caption{\label{block_diagram}Block diagram of SCF method for one pixel: If $X$ has already happened in conjunction with a similar pattern, it is coded in Stage 1. Otherwise, if the color already appeared in the image, it is coded via palette-based coding. If neither case has occurred, it is encoded using residual coding. Finally, the histograms and the pattern list are updated.}
	\end{figure}

\begin{figure}
	 \hfil\includegraphics[scale=0.15]{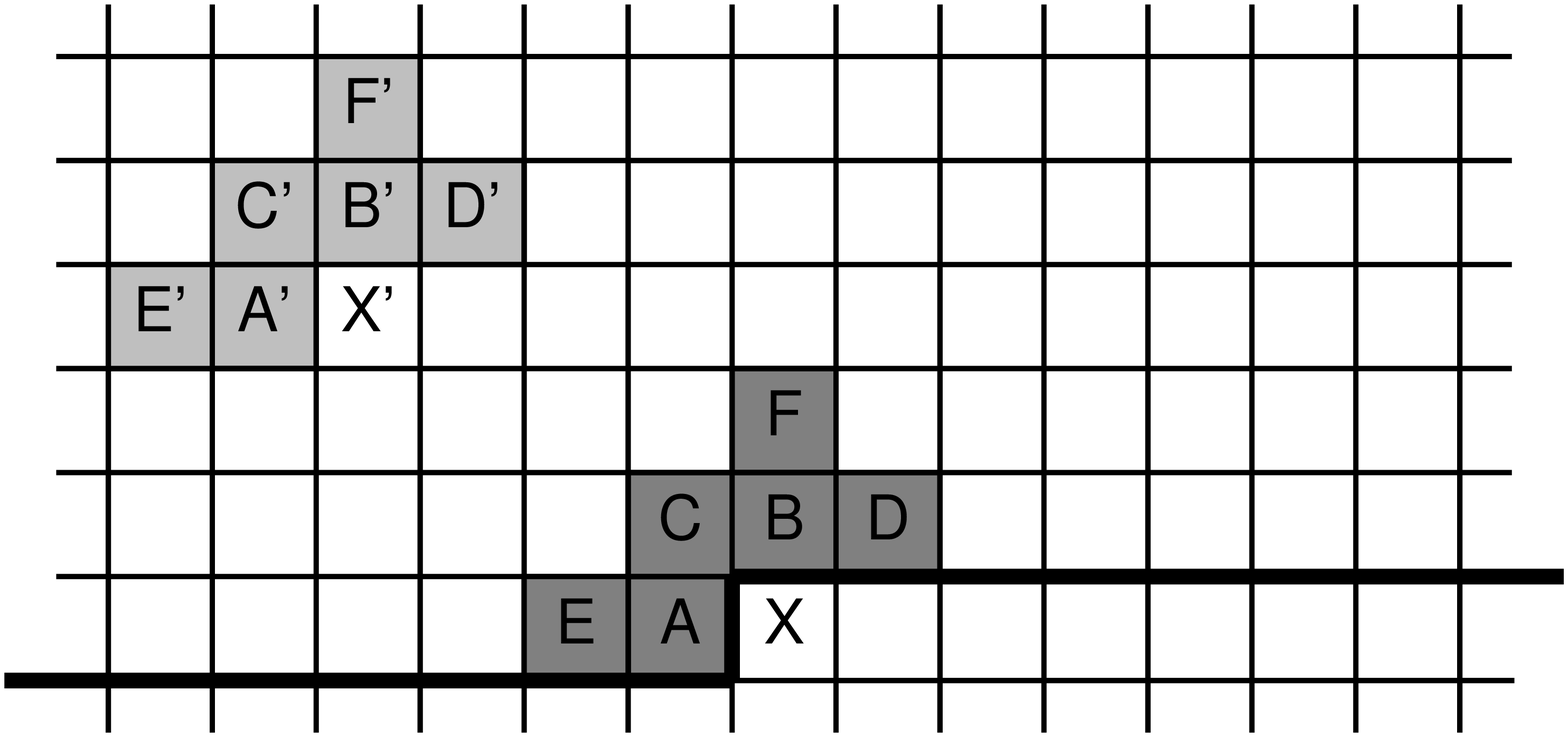}
		\caption{\label{fig_template_matching}Context pattern: If the values in the template (A, B, \dots , F) are similar to (A', B', \dots, F'), then the current value at position X is likely to be similar to the value at X'.}
	\end{figure}

The SCF method is based on the principal of ideal entropy coding. The general processing order is visualized in \Figu{block_diagram}. 
The pixels are scanned in raster-scan order and for each pixel a probability distribution is estimated based on knowledge gained from the already encoded pixels of the image. To this end, for each processed pixel the neighboring pixels in pattern $P = \{A,B,C,D,E,F\}$ are regarded as the context, as illustrated in \Figu{fig_template_matching}.  
For each unique context, the occurrences of the colors that occur in conjunction with that context are counted. When encoding a color pixel (an entire color triple at once), the first step is to search for similar patterns in the already encoded parts of the image.
A pixel-wise similarity measure is used that results in a maximum similarity $s$ for each search. 
The value of $s$ corresponds to a number from 0 to 6 and is equal to the number of pixels in the pattern that were found to be similar. For more information about the similarity criterion, please refer to \cite{Str16b}. Using the associated color histograms of all similar patterns that have maximum $s$, a probability distribution for the current pixel is estimated. Let $n(c|P)$ be the number of occurrences of color $c$ associated with a particular pattern $P$ and the total number of occurrences of pattern $P$ be given as $N_P$. Then the probability of color $c$ given pattern $P$ is estimated as 
	\begin{equation}
		p(c|P) \approx \frac{n(c|P)}{N_P}.
	\end{equation}
If the color of the current pixel is part of the estimated distribution, it can be directly coded using a multi-symbol arithmetic coder, which outputs roughly the corresponding information 
	\begin{equation}
		I(c|P)=-\log_2{p(c|P)}\; \text{[bit]}.
	\end{equation}
Coding using this soft context is from now on called Stage~1 coding. Has the color not happened before in combination with a similar pattern, it is not part of the distribution and thus cannot be encoded this way. Instead, an escape symbol is encoded to signal the transition to Stage 2. Here, a probability distribution is generated based on a color palette. To this end, the occurrences of each color are counted during the encoding process. If the current color has already appeared in the image, it can be encoded in Stage 2. 
However, in case the current pixel has a completely new color, it is not yet part of the palette. An escape symbol signals switching to Stage 3. In Stage 3 (residual coding), the current color is predicted using the extended median adaptive predictor (MAP, \cite{Par04,Bed04}), for each color component separately. The predictor is enhanced by a component-wise adaptation (MAPc, \cite{Str16b}), where the assumption is utilized that directional structures are similar in all components. Depending on the best fit from a set of predictors in the last coded component, the predictor for the next one is chosen. One histogram per component, counting the occurrences of prediction errors, is used to generate a probability distribution for the entropy coding.

In general, SCF compresses screen content very efficiently as long as the number of distinct colors is not too high. Stage 1 can encode pixels at low bitrates of typically $<0.342$\,bits per pixel. 
Stages 2 and 3, however, are less efficient. This means that especially images with mixed content, e.g. screenshots containing large pictorial regions, are less efficient to compress with the SCF method, since they usually contain more patterns and colors and their pixels have to be encoded more often with Stage 2 and 3. Therefore, improving Stage 2 and 3 coding is very important to facilitate the efficiency of the SCF method even for images with mixed content.
	%
\section{Proposed Improvements to Soft Context Formation}
\subsection{Optimization of Residual Error Coding}
\label{stage3}

The residual coding stage is the least efficient stage in the SCF method. Thus, this is a good target for enhancement.
The basic idea of the proposed modification is a pruning of the residual error distribution based on the values of the direct neighbors of the current pixel. In \cite{How93}, the intensity differences between the pixel above and to the left of the current position are established as a prediction context.
A range is identified from the smaller to the larger values of these two pixels. One bit is used to encode whether the current value is within or outside the computed range, and when out of range, one further bit has to be transmitted to signify whether the value is above or below the range. 

The proposed scheme is based on a similar approach. It chooses and trims the probability distributions for the prediction error based on an automatically determined range. 
\begin{algorithm}
	\DontPrintSemicolon 
	\For{pixel $x(i,j)$ in Stage 3} {
		\For{$k$ in color components}{
		$t_k \gets e_{k,\mathrm{max}}/36$ \tcp*{threshold}
		$e_{k}(i,j) \gets x_k(i,j) - \hat{x}_{k,\mathrm{MAP}}(i,j)$ \tcp*{prediction error}
		$r_{k}(i,j) \gets \max \Big($\parbox[t]{.6\linewidth}{$|e_{k}(i-1,j)|,|e_{k}(i-1,j-1)|,$\\ $|e_k(i,j-1)|,|e_{k}(i+1,j-1)|\Big)+1 $\tcp*{range}}\;
			\uIf{$r_{k}(i,j) \leq t_k$} {
				\uIf{$|e_{k}(i,j)| \leq r_{k}(i,j)$} {
				\tcc{Case 1: In-range prediction error}
				Encode `in-range' decision\;
				Encode $e_k(i,j)$ based on `in-range' histogram\;
				Update `in-range' histogram\;
				Increment `in-range' count\;
				}
				\Else {
				\tcc{Case 2: Out of range prediction error}
				Encode `out-of-range' decision\;
				\uIf{$e_k(i,j) \leq 0$} {$e_{k}(i,j) \gets e_{k}(i,j)+r_k(i,j)$\;}
				\Else {$e_{k}(i,j) \gets e_{k}(i,j)-r_k(i,j)-1$\;}
				Encode $e_k(i,j)$ based on `out-of-range' histogram\;
				Update `out-of-range' histogram\;
				Increment `out-of-range' count\;
				}
			}
			\Else{
			\tcc{Case 3: Range is bigger than threshold}
			$e_{k,c}(i,j) \gets x_k(i,j) - \hat{x}_{k,\mathrm{MAPc}}(i,j)$\;
			Encode $e_{k,c}(i,j)$ based on `case 3' histogram\;
			Update `case 3' histogram\;
			}
		}
	}
	\caption{Stage 3 Resdiual Coding}
	\label{algo:stage3coding}
\end{algorithm}
\Algo{algo:stage3coding} describes the coding process. 
Let $i$ be the number of the current column, $j$ be the number of the current row.
Then, the residual errors of component $k$ at position $(i,j)$ using the MAP and MAPc predicted values $\hat{x}_{k,\mathrm{MAP}}(i,j)$ and $\hat{x}_{k,\mathrm{MAPc}}(i,j)$ are given as $e_{k}(i,j) = x_k(i,j) - \hat{x}_{k,\mathrm{MAP}}(i,j)$ and $e_{k,c}(i,j) = x_k(i,j) - \hat{x}_{k,\mathrm{MAPc}}(i,j)$, respectively.
From the errors at adjacent pixel positions, an adaptive range $r_k(i,j)$ is derived:
	\begin{equation}
		\begin{split}
			r_k(i,j) = \max \Big( & |e_k(i-1,j)|,|e_k(i-1,j-1)|, \\
			                      & |e_k(i,j-1)|,|e_k(i+1,j-1)| \Big)+1
														\;.
		\end{split}
	\end{equation}
The addition of 1 accounts for the case that all included prediction errors are zero and additionally loosens the conditions for the decision explained below. 

This adaptive range $r_k(i,j)$ is compared to a threshold $t_k$ which is empirically chosen as 1/36 of the maximum possible absolute prediction error per component
$e_{k,\mathrm{max}}$.
This thresholding ensures that the following scheme is applied only for regions with a reasonably good prediction quality, where it is more likely that the current prediction error is also relatively small. Furthermore, the thresholding excludes large prediction errors from the distribution of in-range errors to prevent a dilution of said distribution. As such, if $r_k(i,j) \le t_k$ holds, it is likely that the current prediction error is limited and a special treatment is initialized. 
The encoder sends a binary decision whether the current absolute error is within the range of the errors in the causal neighborhood ($|e_k(i,j)| \le r_k(i,j)$).
The probabilities of this decision depend on the relation of true and false decisions at former occurrences which are counted during the encoding process.
If the current absolute error fulfils the condition, a distribution based on all `in-range' errors is generated. This distribution can obviously be limited to $[-r_k(i,j), r_k(i,j)]$.
Otherwise, if $|e_k(i,j)| > r_k(i,j)$, a distribution is chosen for probability modeling that results from all `out-of-range' errors. Since no absolute values smaller than or equal to $r_k(i,j)$ can occur, the prediction error in this case is adjusted by removing the unnecessary offset and the histogram is trimmed to $[-e_{k,\mathrm{max}}+r_k(i,j), e_{k,\mathrm{max}}-r_k(i,j)-1]$. Here, MAP is used instead of MAPc, since the more natural image regions with usually smaller and smoother prediction errors, which often fall in the range, are generally better predicted without component-wise adaptation.

The third case applies, if the range $r_k(i,j)$ is bigger than the threshold $t_k$. Then, the whole histogram of prediction errors occurring in this third case is used for probability modelling with the prediction error computed using MAPc. 
Thus, three different prediction error histograms are maintained per component for the three possible cases, and each time a prediction error is coded, the respective histogram is updated by incrementing the corresponding bin.
	%
\subsection{Enhanced Estimation of the Escape-Symbol Probability}
\label{esc2}
	\begin{figure}[tb]
		\centering
		\subfloat[]{\includegraphics[scale=0.8]{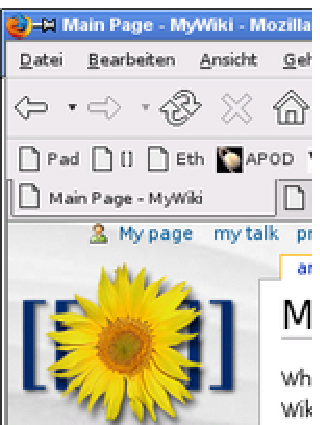}}\hfil
		\subfloat[]{\includegraphics[scale=0.8]{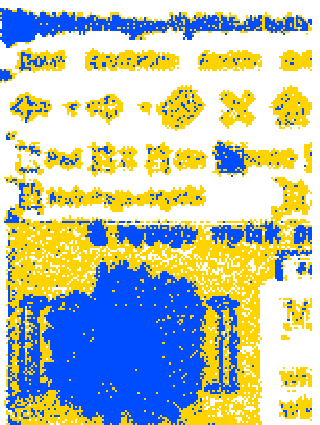}}\hfil
		\caption{\label{fig_esc_array}Example section of screen content image from test set 3: (a) the original RGB image; (b) the positions where the decision for exception handling in Stage 2 is needed. Yellow indicates a known color, while blue indicates the color is new. Pixels marked white are encoded in Stage 1.}
	\end{figure}
A second novel improvement addresses the probability modeling of the escape symbol, which signals a switch from Stage 2 to the residual coding stage. Each time a pixel cannot be encoded in Stage 1, the information is sent whether the current color is already an element of the color palette or not. 
It has been reported in \cite{Str19}, that a conditional estimation of the probability of the escape symbol based on the maximum similarity ($s$, see Section \ref{review}) of the patterns found in Stage 1 is a better model than using the global occurrence. 
The assumption is, that a low similarity count (i.e., the current pattern is not at all similar to other patterns previously found in the image) implies a high probability of a new color. However, it is still relatively likely that a pattern is new, yet the color already known, resulting in a moderate probability estimate for small similarity scores. Instead, it can be seen that the occurrence of new colors is locally dependent. An example image and the positions where the decision has to be sent whether a color is new or not can be seen in \Figu{fig_esc_array}. 
	
Therefore, it is proposed to model the probability of the escape symbol in Stage 2 based on the occurrence of novel colors in the neighborhood, utilizing the positions $A,B,C,D,E$ and $F$, similar to \cite{Str20}. In total, 64 different contexts $\mathrm{ctx}$ are taken into account, for each possible variation of the binary decisions, whether positions $A,B,C,D,E$ and $F$ contained previously unseen colors or not. By counting the occurrence of a new color at position $X$ given a certain context $\mathrm{ctx}$ for each pixel not encoded in Stage 1 as $n(\mathrm{ESC}=\mbox{true}|\mathrm{ctx})$, the probability of the escape symbol is estimated as
	\begin{equation}
		p(\mathrm{ESC}=\mbox{true}) \approx \frac{n(\mathrm{ESC}=\mbox{true}|\mathrm{ctx})}{N_{\mathrm{ctx}}}
	\end{equation}
with $N_{\mathrm{ctx}}$ the total count of the occurrence of $\mathrm{ctx}$. Regular downscaling of these counts enables adaptation to changing image content.
Using these contexts to model the likelihood takes into account two aspects simultaneously: First, the local frequency of new colors, i.e., the information whether this is a region with many new colors, and second, the position of new colors with respect to the current pixel, by considering whether, for example, only the pixels at more distant positions $E$ or $F$ contain new colors, or the immediate neighbors $A$ and $B$. 
\section{Evaluation}
\label{results}
For evaluation, SCF with the proposed modifications is compared to the SCF version from \cite{Str20} as well as other state-of-the-art methods. To this end, six different RGB test sets are utilized. Test sets 1 to 4 correspond to those used in \cite{Str20} for comparability. The images in test set 1 contain up to 7372 different colors. Test set 2 are still images taken from the HEVC test sequences. A mixture of images with varying properties composes test set 3. Test set 4 consists of images from the SIQAD database \cite{Hua15,SIQAD}.
36 images from the SCID database \cite{Zha17,ESIM17,SCID} form test set 5. Finally, test set 6 contains further 86 diverse screen content images, leading to a total amount of 306 test images. 

In \Tabl{results_all}, the absolute file sizes after lossless compression using FLIF \cite{Sne16}, HEVC \cite{HEVCSCC} (HM 16.21 + SCM 8.8)  and SCF with and without the proposed modifications are listed for each dataset. Additionally, the relative performance is shown underneath. As one can see, the original SCF coding scheme from \cite{Str20} already surpasses the other compression methods for each dataset in sum. The proposed modifications gain further $0.5\,\%$ bit savings for set 1, around $2\,\%$ for sets 2, 3 and 6, $4.9\,\%$ for data set 4, and $3.3\,\%$ for test set 5.
\begin{table}[t]
\captionsetup{font=footnotesize}
	\caption{\label{results_all}Comparison of compression performance of the proposed SCF method with state-of-the-art methods. The table lists file sizes in bytes as well as relative file sizes in comparison to the proposed method.}
	\hfil
	{\setlength{\tabcolsep}{2pt}
		\begin{tabular}{|l|c|c|c|c|cc|}
\hline
              &   Num.  &   Num.   			&											&         						& \multicolumn{2}{c|}{SCF}  												\\
              &   of    &   of     			&											& HM-16.21						& Original 						& Proposed 										\\
              &  images &  colors 			& FLIF								& SCM-8.8 						& Ref. \cite{Str20}		&         										\\ \hline \hline
Test set 1    & 67			& \multicolumn{1}{l|}{2 …}
																				& 2844684           	& 2566530           	& 1922840           	& \textbf{1913761}          	\\
Percentage 		&         & 7372          & \textit{148.6\,\%}  & \textit{134.1\,\%}  & \textit{100.5\,\%}  & \textit{\textbf{100.0\,\%}} \\ \hline
Test set 2    & 14			& \multicolumn{1}{l|}{505 …}
																				& 3127442           	& 2981071           	& 2693302           	& \textbf{2623851}          	\\
Percentage 		&         & 67477         & \textit{119.2\,\%}  & \textit{113.6\,\%}  & \textit{102.6\,\%}  & \textit{\textbf{100.0\,\%}} \\ \hline
Test set 3    & 83			& \multicolumn{1}{l|}{7 …}
																				& 9821855           	& 8082341           	& 6939066           	& \textbf{6788148}          	\\
Percentage 		&         & 88751         & \textit{144.7\,\%}  & \textit{119.1\,\%}  & \textit{102.2\,\%}  & \textit{\textbf{100.0\,\%}} \\ \hline
Test set 4    & 20			& \multicolumn{1}{l|}{6026 …}
																				& 5132256           	& 4997400           	& 4936375           	& \textbf{4704624}          	\\
Percentage 		&         & 162170        & \textit{109.1\,\%}  & \textit{106.2\,\%}  & \textit{104.9\,\%}  & \textit{\textbf{100.0\,\%}} \\ \hline
Test set 5    & 36			& \multicolumn{1}{l|}{22 …}
																				& 14902270          	& 14643675          	& 14471624          	& \textbf{14004404}         	\\
Percentage 		&         & 340656        & \textit{106.4\,\%}  & \textit{104.6\,\%}  & \textit{103.3\,\%}  & \textit{\textbf{100.0\,\%}} \\ \hline
Test set 6    & 86			& \multicolumn{1}{l|}{2 …}
																				& 13268411         		& 12695705     	    	& 11733092          	& \textbf{11451696}         	\\
Percentage 		&         & 438637				& \textit{115.9\,\%}  & \textit{110.9\,\%}  & \textit{102.5\,\%}  & \textit{\textbf{100.0\,\%}} \\ \hline \hline
Total					& 306			&								& 49096918						& 45966722						& 42696299							& \textbf{41486484}						\\ 
Percentage		&					&								& \textit{118.3\,\%} 	& \textit{110.8\,\%}	& \textit{102.9\,\%}	& \textit{\textbf{100.0\,\%}} \\ \hline
\end{tabular}

	}
\end{table}

To shed some more light on the savings for each proposed modification separately, \Tabl{results_mod} contains the file size savings for the SCF method only with the modification for the escape symbol in Stage 2, as well as with both proposed modifications implemented. 
\begin{table}[t]
\captionsetup{font=footnotesize}
	\caption{\label{results_mod}Investigation of the effect on the two proposed modifications. The table lists file sizes of the compressed images in bytes.}
	\hfil \begin{tabular}{|c|c|c|c|c|}
\hline
   Percentage of 		& Num. 	& \multicolumn{3}{c|}{SCF} 														\\
 unique colors  		&   of   	&  Original          	& With stage 2 	& With both 		\\
  per image      		& images 	&  Ref. \cite{Str20} 	& modification 	& modifications \\ \hline \hline
$\leq$ 3\,\%        & 205    	& 10332075            & 10284912      & 10222359      \\
Percentage					& 				& 101.1\,\% 					& 100.6\,\% 		& 100.0\,\%  		\\ \hline
$\leq$ 7\,\%        & 47     	& 9632475             & 9572655       & 9415846       \\
Percentage	 				& 				& 102.3\,\% 					& 101.7\,\% 		& 100.0\,\%  		\\ \hline
$\leq$ 17\,\%       & 32    	& 11364574            & 11268737      & 10980798      \\
Percentage 					& 				& 103.5\,\% 					& 102.6\,\% 		& 100.0\,\%   	\\ \hline
$>$ 17\,\%       		& 22     	& 11367175            & 11268671      & 10867481      \\
Percentage 					& 				& 104.6\,\% 					& 103.7\,\% 		& 100.0\,\%   	\\ \hline 
\end{tabular}

\end{table}
The results are separated into sets depending on the amount of unique colors per image, such, that each class has roughly the same amount of bytes when compressed. It is evident, that the more different colors an image contains the stronger the influence of the proposed enhancements. For images with a percentage of unique colors less than $3\,\%$, the modification of the residual coding stage saves $0.6\,\%$ while for images where more than $17\,\%$ of the colors are unique $3.7\,\%$ file size can be saved. Additionally, it can be seen, that the proportion of the overall savings gained by the residual coding modification with respect to the total savings rises with the percentage of different colors per image. This can easily be explained, since the proposed modification to Stage 3 only effects the bitrate when a completely new color is encoded, whereas the escape symbol modification takes effect whenever a pixel could simply not be encoded in Stage 1.\\
The increase in coding time introduced by the modifications is negligible in comparison to the total computation time.
	%
\section{Summary}
	%
SCF is an efficient compression scheme for screen content images with few colors, but less so for images containing many colors and different patterns, where fewer pixels can be encoded based on the soft context model. Instead, palette-based coding or residual encoding is performed in these cases. When estimating the probability of a new color, it could be shown that taking the local frequencies and the relative positioning into account  improves the results compared to conditioning based on the similarity of the best similar context found. Furthermore, adaptive pruning of the prediction error histograms significantly improved the residual coding stage, which has a greater effect the more colors an image contains.
\section*{Acknowledgement}
This work has been funded by the Deutsche Forschungsgemeinschaft (DFG, German Research Foundation) - 438221930. 
\balance
\bibliographystyle{IEEEtran}
\bibliography{literature}

	%
\vspace{12pt}

\end{document}